# Three-Class Emotion Classification for Audiovisual Scenes Based on Ensemble Learning Scheme

XIANGRUI XIONG, GMIT Laboratory of Xihua University, Chengdu 611730, China.
ZHOU ZHOU, Beibu Gulf University, Qinzhou 535011, China.
GUOCAI NONG, Beibu Gulf University, Qinzhou 535011, China.
JUNLIN DENG, Beibu Gulf University, Qinzhou 535011, China.
NING WU[*1], Beibu Gulf University, Qinzhou 535011, China.

Emotion recognition plays a pivotal role in enhancing human-computer interaction, particularly in movie recommendation systems where understanding emotional content is essential. While multimodal approaches combining audio and video have demonstrated effectiveness, their reliance on high-performance graphical computing limits deployment on resource-constrained devices such as personal computers or home audiovisual systems. To address this limitation, this study proposes a novel audio-only ensemble learning framework capable of classifying movie scenes into three emotional categories: Good, Neutral, and Bad. The model integrates ten support vector machines and six neural networks within a stacking ensemble architecture to enhance classification performance. A tailored data preprocessing pipeline, including feature extraction, outlier handling, and feature engineering, is designed to optimize emotional information from audio inputs. Experiments on a simulated dataset achieve 67% accuracy, while a real-world dataset collected from 15 diverse films yields an impressive 86% accuracy. These results underscore the potential of audio-based, lightweight emotion recognition methods for broader consumer-level applications, offering both computational efficiency and robust classification capabilities.

CCS CONCEPTS：**Artificial Intelligence→Machine Learning→Ensemble Learning**

## 1 INTRODUCTION

In recent years, the rise of short video platforms powered by intelligent recommendation algorithms has significantly increased user engagement on smartphones. This surge in content consumption has amplified the demand for efficient video editing tools, particularly those capable of identifying emotionally relevant clips. Accurate emotion classification is essential for applications such as content filtering, personalized viewing experiences, and adaptive audiovisual effects. For example, certain movie scenes—especially those with intense or unpleasant emotional content—may be unsuitable for family viewing environments. Furthermore, intelligent systems that dynamically adjust sound effects based on emotional transitions require reliable and real-time emotion recognition from audiovisual data. To ensure practical applicability in household contexts, such recognition algorithms must operate efficiently on consumer-grade devices with limited computational resources, rather than relying on high-end professional workstations.

---

* Corresponding author
Authors' addresses: n.wu@bbgu.edu.cn

Emotion identification in movie clips is also critical for content retrieval, automated summarization, and sentiment-aware recommendation. Existing research has explored three primary approaches to emotion recognition: video-based machine vision techniques [1], audio-based methods [2], and multimodal frameworks that combine both modalities [3]. A more experimental line of work includes EEG-based emotion detection during movie viewing [4]. However, each of these methods faces significant challenges. Visual and multimodal approaches require substantial graphics processing capabilities, limiting their deployment on mainstream devices. EEG-based systems, while offering high accuracy, are constrained by the need for specialized hardware, making them impractical for everyday use.

Audio-based techniques offer a promising alternative due to their lower computational demands and broader hardware compatibility [2]. Nonetheless, most existing audio emotion classification methods focus on binary tasks (e.g., shock vs. non-shock), which are inadequate for real-world applications. There remains a need to extend these methods to handle more complex, fine-grained emotional categorizations.

This study addresses this gap, and a novel audio-only ensemble learning approach is proposed to classify movie segments into three emotion categories such as Good, Neutral, and Bad. The model architecture employs a stacking strategy that integrates ten support vector machines (SVMs) and six neural networks as base learners, followed by a meta-learner to refine the classification output. This layered ensemble framework leverages the strengths of diverse learning models to enhance emotion recognition accuracy and robustness.

Experimental evaluations on a simulated movie audio emotion dataset are carried out and tests on a carefully curated real-world movie dataset demonstrate the feasibility and effectiveness of the proposed approach in realistic application scenarios. In parallel, we developed a comprehensive data preprocessing pipeline—including segmentation, feature extraction, outlier detection, and feature engineering—to ensure that emotionally salient information is optimally captured and utilized by the model.

The ensemble learning framework offers a practical, scalable, and high-performing solution for three-class emotion recognition in movie audio. It not only surpasses traditional binary classification methods but also addresses key limitations associated with vision-based and EEG-driven approaches.

The remainder of this paper is organized as follows. Section 2 reviews existing literature on emotion classification in multimedia. Section 3 introduces the proposed ensemble learning architecture. Section 4 discusses model optimization strategies including grid search and nested cross-validation. Section 5 presents experiments on simulated datasets to preliminarily validate the model. Section 6 details evaluations on a real-world movie audio emotion dataset. Finally, Section 7 concludes the paper with a summary of findings and future directions.

## 2  RELATED WORKS

Various methods have been proposed to address the challenge of emotion recognition in movie content, each with distinct advantages and limitations. One prominent line of research involves real-time emotion recognition using electroencephalogram (EEG) signals. A notable EEG-based approach achieves classification accuracies of 86.63% for positive and negative emotions, and 92.26% for neutral clips characterized by high arousal and valence levels [4]. Despite its promising real-time capabilities, this method is heavily reliant on specialized EEG detection equipment, confining its applicability to controlled laboratory environments and limiting its feasibility for consumer-oriented scenarios.

Another widely explored approach is facial expression analysis, which infers emotional states by interpreting visual cues in facial movements. While this method reaches an impressive 90% accuracy for negative emotions, its performance drops significantly for neutral and positive emotions, achieving only 50% accuracy [1]. Furthermore, the computational demands of facial analysis—requiring high-performance graphics hardware—restrict its use in everyday home devices.

In contrast, audio elements such as music and speech offer a more practical modality for emotion recognition. As an integral component of cinematic expression, music carries substantial emotional information [6, 7]. Parallel studies in speech emotion recognition suggest that vocal attributes can effectively convey affective states, underscoring the emotional significance embedded in audio streams of film content [8].

Multimodal methods that combine audio and visual data have also shown potential in emotion recognition tasks. For example, an algorithm developed using the eNTERFACE '05 Audio-Visual Emotion Database demonstrated 75% accuracy in recognizing six emotional states through combined modalities [3, 9]. However, this dataset is limited to general voice signals and lacks specialized movie emotion data, casting uncertainty on its effectiveness in cinematic contexts.

Audio-only methods provide a computationally efficient alternative. Jin et al. proposed an approach that analyzes audio segments using a 7-second window combined with low-pass filtering, achieving 91.5% accuracy in classifying shocking versus non-shocking emotional scenes [2]. This high performance is largely attributed to the pronounced acoustic signatures of startling scenes—often represented by bass-heavy sound effects. While this binary classification is useful, it highlights the need for further research into multi-class emotion recognition.

Building on this foundation, the present study proposes an audio-only ensemble learning framework that integrates SVM and neural networks (NNs) to classify movie segments into three emotional categories: Good, Neutral, and Bad [10]. The model's reliance solely on audio input reduces computational overhead and training time, making it suitable for deployment on resource-constrained platforms such as personal computers, set-top boxes, advertisement displays, industrial control systems, and mobile devices. This research contributes to the field by expanding beyond binary classification and offering a lightweight, scalable solution for real-world emotional content recognition.

## 3  THE ENSEMBLE LEARNING SCHEME

Ensemble learning is a powerful machine learning paradigm that integrates multiple weak learners into a single strong learner to improve overall predictive performance. Typically, ensemble models are structured in two layers: the base learners, which extract diverse features from the data, and the meta learner, which consolidates their outputs. The interaction between these layers varies according to the type of ensemble strategy adopted—most notably bagging, boosting, and stacking. In bagging and boosting, the meta learner is trained on the base learners' outputs derived from the training set. In contrast, stacking utilizes outputs from a validation set to train the meta learner, thereby reducing overfitting and improving generalization performance [11]. In this study, we adopt the stacking approach due to its effectiveness in minimizing generalization error and enhancing classification robustness [12].

Our proposed model employs a heterogeneous set of base learners to capture complementary features from the input data. Specifically, the base layer consists of three backpropagation neural networks (BPNNs), nine SVM, and three one-dimensional convolutional neural networks (1D-CNNs), as illustrated in Figure 1. These diverse models extract a rich set of high-level features from the audio input samples. The outputs from all base learners are then concatenated and used as new input features for a support vector machine, which serves as the meta learner within the stacking ensemble framework to perform the final emotion classification.

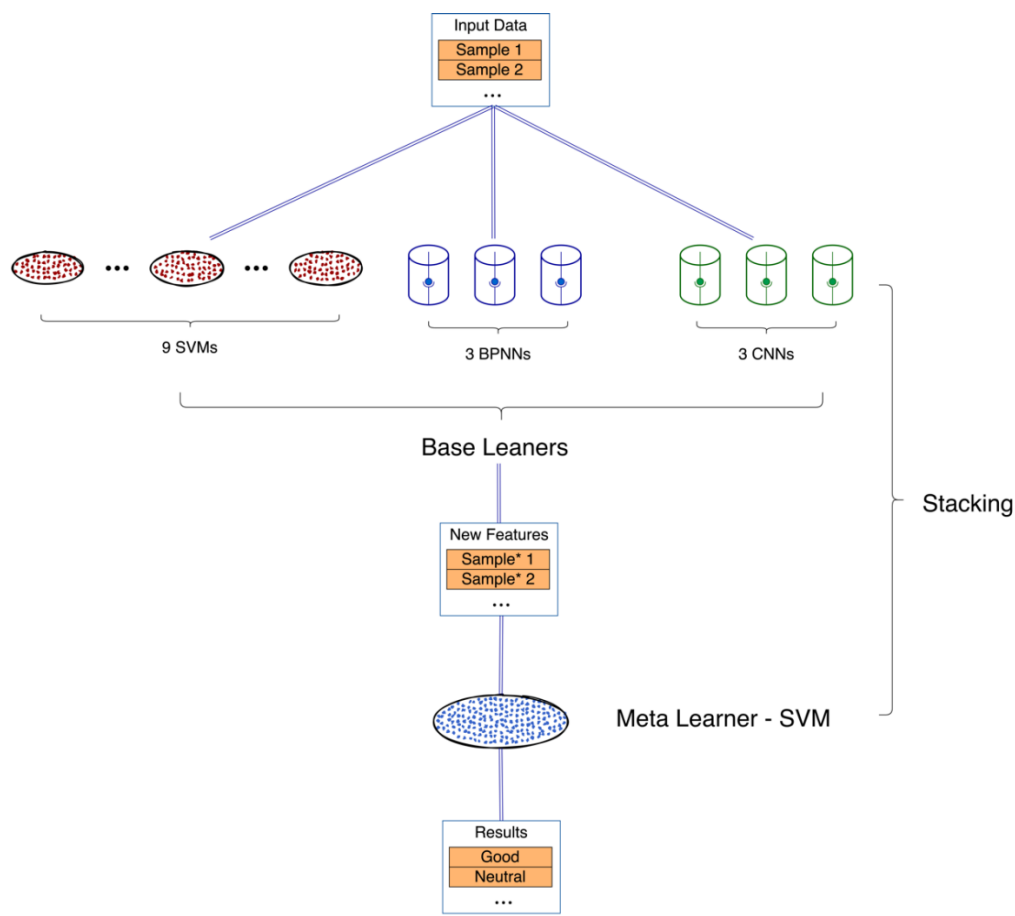

Figure 1. The ensemble schemes.

In order to capture a broad range of high-level emotional features, deliberate variability among the base learners can be employed. Although the base SVM learners exhibit limited sensitivity to neutral emotions and struggle with accurately identifying negative emotions, they perform exceptionally well in recognizing audio patterns associated with the Good emotion category. In contrast, neural networks tend to underperform in classifying negative emotions and offer only moderate compensation for the system's insensitivity to neutral expressions. Nevertheless, by leveraging the complementary strengths of these diverse models, the ensemble of base learners is capable of generating informative features related to Bad emotions—features that are not effectively captured by any single model in isolation. These emergent features are sufficiently distinct for the meta learner to recognize new instances of Bad emotions.

The challenge of accurately identifying Bad emotion classes cannot be effectively addressed using simple voting strategies, which are commonly applied in ensemble learning. This limitation arises because all base learners consistently misclassify Bad samples as belonging to other categories, rendering majority voting ineffective. In contrast, the meta learner, implemented using SVMs, is capable of extracting emotion-related patterns embedded in the new representations generated by the base learners, thereby constructing more sophisticated decision boundaries for improved classification performance.

To achieve optimal performance, the Grid Search method is employed iteratively to fine-tune the SVM hyperparameters within the ensemble learning framework. In each iteration, the hyperparameters are updated until the best-performing combination is identified. In parallel, neural networks utilize an Early Stopping strategy during training; at each early stopping point, a new neural network is generated through repeated applications of the Early Stop procedure.

### 3.1 Design base learners

*3.1.1 Design SVMs*

The kernel function is a critical component of SVMs, as it enables the mapping of non-linearly separable samples from a low-dimensional space to a higher-dimensional feature space, where more effective classification becomes possible. Among various kernel types, the Radial Basis Function (RBF) kernel has demonstrated strong performance in SVM applications. Accordingly, the SVMs in this ensemble learning framework adopt the RBF kernel to measure the similarity between input samples [13–15]. The RBF kernel can be given as,

$$K(x^{(i)}, x^{(j)}) = \exp\left\{-\frac{\|x^{(i)} - x^{(j)}\|^2}{2\sigma^2}\right\} \tag{1}$$

When samples are passed through the Radial Basis Function (RBF) kernel, an output value close to 1 indicates high similarity between the two samples—suggesting that they likely belong to the same class—while a value approaching 0 suggests that the samples are from different classes. Although the RBF kernel's formulation does not explicitly reveal the feature mapping process, it implicitly performs an inner product operation in a transformed feature space, thereby projecting the original low-dimensional data into a high-dimensional space. This transformation enables linear separation of otherwise non-linearly separable data [16].

The RBF kernel includes a key parameter that controls the width of the kernel function, which in turn affects the curvature and complexity of the decision boundary as well as the model's generalization capability. In this study, the Support Vector Classifier (SVC) implementation from the Scikit-learn library is employed, where the kernel parameter is denoted as $\gamma$. A lower $\gamma$ value yields a smoother decision boundary, promoting wider margins and reducing the risk of overfitting. In contrast, a higher $\gamma$ value produces a more intricate decision boundary, emphasizing local data structure and potentially reducing training errors at the cost of generalization [17].

Another crucial parameter in SVC is the penalty term C, which governs the trade-off between maximizing the margin and minimizing classification errors. A higher C value imposes stricter penalties for misclassifications, thereby reducing the tolerance for violations within the margin and potentially overfitting the training data. Conversely, a lower C value allows greater flexibility, which can help prevent overfitting by tolerating some degree of error. The specific selection of the $\gamma$ and C parameters will be detailed in the experimental procedures described in Section 7.

*3.1.2 The design of 1D-CNN*

Convolutional Neural Networks (CNNs) primarily consist of convolutional and hidden layers. Unlike traditional machine learning approaches, CNNs are capable of automatically extracting relevant features from input data, often leading to superior performance [18, 19]. Since audio signals are inherently one-dimensional, a one-dimensional CNN is well-suited for their processing. Accordingly, a base learner utilizing a one-dimensional CNN architecture is constructed, as illustrated in Figure 2.

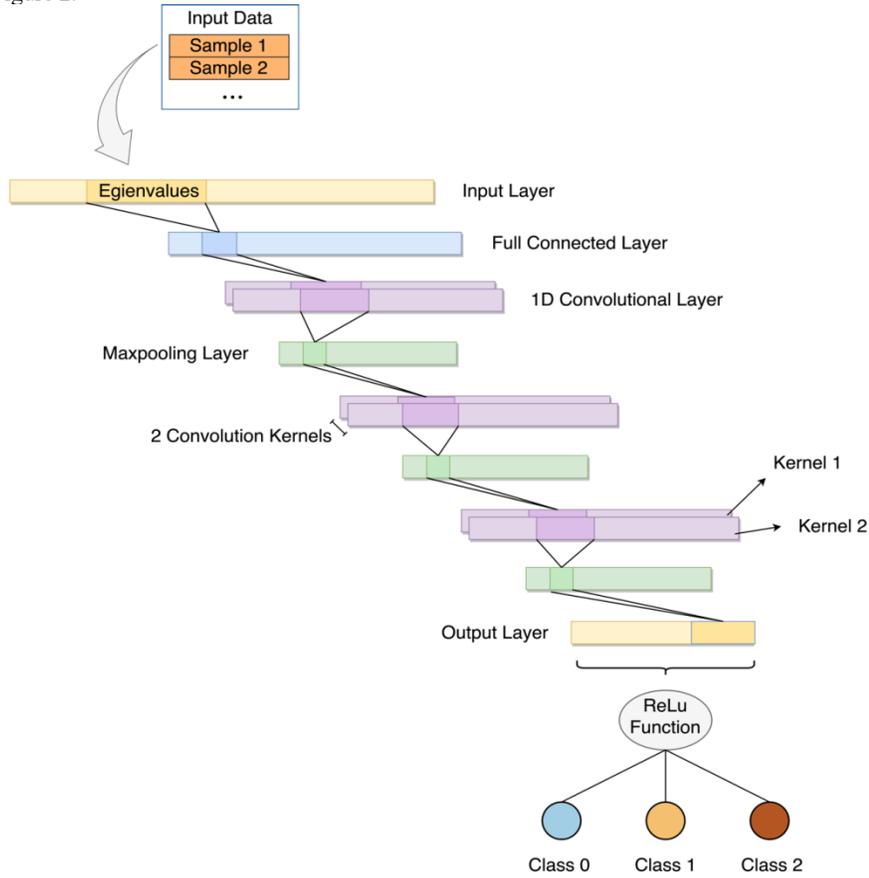

Figure 2. The structure of 1D-CNN.

This base learner employs three 1D convolutional layers to extract features from the input data. For illustrative purposes, only two convolutional kernels are depicted in the figure; however, the actual number of kernels varies across the layers. Each convolutional layer is followed by a max-pooling layer, which serves as a downsampling technique to reduce overfitting and computational complexity. By decreasing the dimensionality of the feature maps, the max-pooling layer lowers the computational load for subsequent layers, thereby improving processing efficiency [20]. The reduction in feature map resolution also enables the following convolutional layers to focus more effectively on extracting representative global features, rather than overemphasizing local details.

To further enhance computational efficiency, a fully connected layer with reduced dimensionality is placed directly after the input layer. This layer contains 69 neurons, which receive 138 signals from the input layer and output 69 signals to the first convolutional layer. Functionally, it performs an initial transformation that halves the data volume, reducing the input dimensionality before feature extraction begins. Furthermore, dropout is applied to selected neural layers during training. This regularization technique randomly disables a subset of neurons, preventing overreliance on specific pathways and enhancing the overall robustness and generalization ability of the network [21]. The detailed hyperparameters for this 1D-CNN architecture are summarized in Table 1.

Table 1. The parameter setting of the 1D-CNN structure.

| Layer Type | Hyperparameters |
|---|---|
| Full Connected | in_features = 138 |
|  | out_features = 69 |
|  | dropout rate = 20% |
| 1D CNN | in_channels = 1 |
|  | out_channels = 26 |
|  | kernel_size = 2 |
|  | dropout rate = 30% |
| Maxpooling | kernel_size = 2 |
| 1D CNN | in_channels = 26 |
|  | out_channels = 16 |
|  | kernel_size = 2 |
|  | dropout rate = 40% |
| Maxpooling | kernel_size = 2 |
| 1D CNN | in_channels = 16 |
|  | out_channels = 16 |
|  | kernel_size = 2 |
|  | dropout rate = 10% |
| Maxpooling | kernel_size = 2 |
| Full Connected | in_features = 120 |
|  | out_features = 3 |
|  | dropout rate = 20% |
| ReLU Function | None |

*3.1.3 The design of BPNN*

The Backpropagation Neural Network (BPNN) is a widely adopted architecture across various research domains [22], capable of dynamically adjusting its internal weights based on the error between the network's predicted output and the actual target output, iteratively minimizing this difference until a satisfactory level of accuracy is achieved [23]. Table 2 presents the specific hyperparameters used for the BPNN, highlighting that layers other than the output layer contain a relatively large number of neurons.

Table 2. Parameter setting of the BPNN structure.

| Layer Type | Hyperparameters |
|---|---|
| Full Connected | in_features = 138 |
|  | out_features = 69 |
|  | dropout rate = 20% |
| Hidden Full Connected | in_features = 69 |
|  | out_features = 41 |
|  | dropout rate = 20% |
| Maxpooling | kernel_size = 2 |
| Hidden Full Connected | in_features = 20 |
|  | out_features = 10 |
|  | dropout rate = 20% |
| Hidden Full Connected | in_features = 10 |
|  | out_features = 5 |
|  | dropout rate = 10% |
| Full Connected | in_features = 5 |
|  | out_features = 3 |
|  | dropout rate = None |
| ReLU Function | None |

The hidden layers of this BPNN consist of three fully connected layers, with a max pooling layer inserted after the first layer. This configuration is based on the assumption that the network's structural complexity is relatively low, and thus it is not necessary to apply a pooling layer after every dense layer—helping to reduce overfitting and lower computational cost.

The output layer employs a Rectified Linear Unit (ReLU) activation function, similar to the design commonly seen in Convolutional Neural Networks (CNNs). The ReLU function allows non-negative linear outputs for positive inputs while setting negative inputs to zero [24, 25]. Strategically placing the ReLU activation in the output layer helps mitigate the well-known issue of "dead" neurons—neurons that consistently output zero—by introducing a form of implicit, randomized pruning. This design enhances the sparsity and robustness of the network, ultimately contributing to improved generalization. Overall, this architecture significantly enhances the model's generalization performance, making it well-suited for complex classification tasks with limited computational overhead.

### 3.2 The design of meta learner

The primary objective of the meta learner is to enhance the overall performance of the ensemble model, particularly in terms of prediction accuracy and generalization ability. A meta learner can be implemented using various machine learning algorithms, including SVM, Random Forests, Decision Trees, and Neural Networks [26]. In this study, considering that the outputs of the base learners serve as high-level representations of emotional features extracted from audio samples, and drawing upon the successful application of SVMs in audio emotion recognition, we adopt an SVM with a Radial Basis Function (RBF) kernel as the meta learner in our proposed ensemble learning framework.

As illustrated in Figure 1, the meta learner treats the outputs of each base learner as new feature inputs and learns to integrate them effectively [12]. This process constitutes a higher-level and more holistic fusion of the predictions from individual base learners, allowing the meta learner to model inter-learner relationships and correct base-level misclassifications. The result is a more robust and generalizable ensemble prediction outcome.

Furthermore, because stacking requires the meta learner to be trained on the outputs derived from the base learners' validation sets—which are often limited in size—it is crucial to adopt appropriate validation strategies. Techniques such as leave-one-out cross-validation may offer greater reliability in this context, ensuring that the performance of the meta learner is accurately and fairly evaluated.

## 4 THE OPTIMIZATION METHODS

### 4.1 Optimization with the Grid Search method

Grid Search, originally designed as an exhaustive search over a predefined subset of the hyperparameter space, systematically explores all possible parameter combinations to identify the optimal configuration [27]. The execution flow of the Grid Search algorithm is illustrated in Figure 4. In this process, new learners are trained sequentially, each corresponding to a unique combination of parameters within the specified search space. K-fold cross-validation is employed to evaluate the performance of each learner under different parameter settings, and the configuration yielding the highest performance is selected as the optimal set of hyperparameters [17, 27, 28].

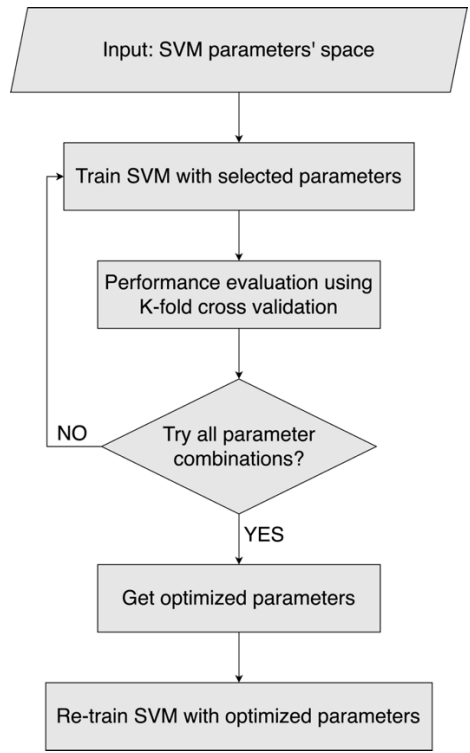

Figure 3. The flowchart of the Grid Search method.

For this study, Grid Search is utilized to determine the best combinations of the regularization parameter C and the kernel parameter γ for each of the nine base SVM learners as well as the SVM meta learner. As shown in Figure 5, model performance across different hyperparameter combinations is evaluated using accuracy as the performance metric. To facilitate visualization, the hyperparameter space is represented as a two-dimensional plane, where the x-axis corresponds to values of C, the y-axis to values of γ, and accuracy is represented along the vertical axis perpendicular to the plane. For instance, the point (1,1) represents C=1 and γ=1.

The topographic map in Figure 5 illustrates the distribution of accuracy values across the hyperparameter space. Brighter yellow regions indicate higher accuracy, while darker or cooler tones represent lower accuracy. This visual representation helps identify the region within the hyperparameter space that yields optimal model performance.

To efficiently converge on the optimal point, the Grid Search is conducted iteratively. In the first iteration, the algorithm explores a broad range of hyperparameter values to identify a region near the global optimum. Based on these initial results, the search range is subsequently narrowed in the second iteration, centering around the best-performing parameters identified previously. This refined search yields a more precise estimate of the optimal hyperparameter combination. The process is repeated until the search converges sufficiently close to the optimal point, ensuring that the proposed model achieves both high accuracy and robust generalization.

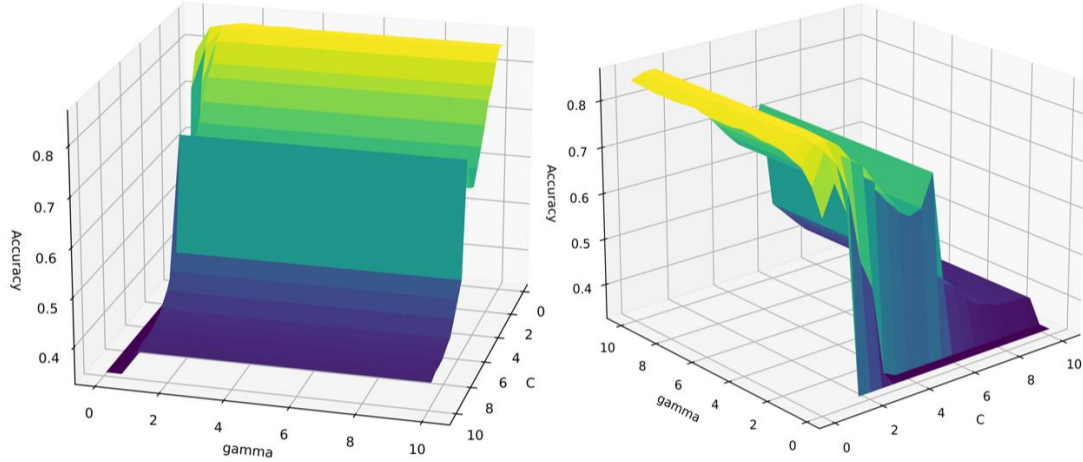

Figure 4．Grid Search Visualization.

**4.2 Nested cross-validationt**

Typically, a subset of the training set is partitioned into a validation set, which is solely used to validate the learner's performance during the training process and does not participate in model training. This could be a problem: bias may develop while evaluating the model. One reason for these phenomena is because the best parameters discovered using a subset of the training set may not be the best parameters throughout the entire training set. Nested cross-validation was introduced as a solution to this problem. Nested cross-validation consists of two layers: outer cross-validation and inner cross-validation.The training set acquired by initially dividing the dataset is stored as the original training set, on which outer layer cross-validation is performed by dividing it into K1 folds, with K1-1 folds serving as the outer training set and the remaining 1 fold serving as the outer validation set. Inner layer cross-validation is performed again on the outer layer training set, which is divided into K2 folds, with K2-1 folds serving as the inner layer training set and the remaining 1 fold as the inner layer validation set [29]. Figure 6 depicts this process. Grid search needs the usage of K folds cross-validation, hence the cross-validation it contains is employed as inner cross-validation in this study. The inner cross-validation determines the best possible combination of hyperparameters, while the outer cross-validation tests that set of hyperparameters, providing an unbiased performance evaluation report for the model with that parameter [29, 30]. This implies that the layered cross-validation-based lattice search algorithm must create many models of K1 * K2 * hyperparameter permutations in order to complete the parameter space search. Because of computational limitations, only five of the nine SVMs are chosen to use nested cross-validation. Such a validation strategy is widely utilized in machine learning research to assure fair evaluation and prevent model overfitting [31].

Typically, a portion of the training set is partitioned as a validation set, which is used exclusively to assess the learner's performance during training and does not participate in model fitting. However, this approach introduces potential bias, as the selected hyperparameters may not generalize well across the full training set. One of the common issues is that the optimal parameters identified on a subset may not represent the true optimal parameters for the entire dataset. To address this limitation, nested cross-validation has been proposed as a robust evaluation strategy, which incorporates two levels of cross-validation: an outer loop for model evaluation and an inner loop for hyperparameter tuning. First, the dataset is divided into a training set and a test set. The training set is then subjected to outer cross-validation, typically by partitioning it into K1 folds. Of these, K1−1 folds are used for training, while the remaining fold serves as the outer validation set. Within each outer fold, inner cross-validation is conducted by further dividing the training data into K2 folds: K2−1 folds are used for model training, and the remaining fold is used for validation. This process is illustrated in Figure 6.

In this study, since Grid Search requires cross-validation to assess model performance under different hyperparameter combinations, it is embedded within the inner loop of the nested structure. The inner cross-validation identifies the optimal hyperparameter set, while the outer cross-validation evaluates the model's performance using those parameters, thus yielding an unbiased performance estimate [29, 30]. This nested strategy enhances reliability by mitigating the risk of overfitting to the validation set.

Implementing nested cross-validation requires training a total of K1×K2×N, where N is the number of hyperparameter combinations, which can be computationally intensive. Due to such limitations, nested cross-validation was applied to only five of the nine SVM base learners in this study. Nevertheless, this technique remains widely adopted in machine learning research as a gold standard for fair performance evaluation and model robustness assurance [31].

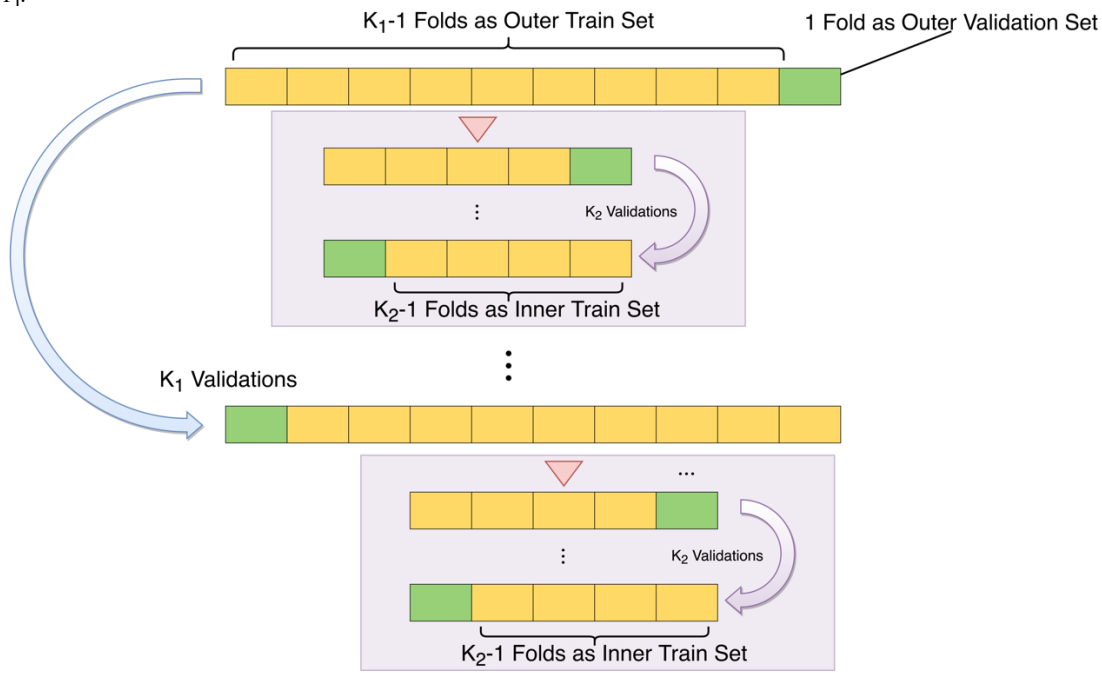

Figure 5．Showing the illustration of nested cross-validation

## 4.3 Leave-one-out cross validation

The stacking ensemble technique requires that the meta learner's training data be derived from the outputs of the base learners on their respective validation sets. However, since validation sets typically comprise only a small portion of the entire dataset, this presents a challenge for the meta learner in effectively dividing training and validation data. To address this, the leave-one-out cross-validation (LOOCV) method is adopted to evaluate the performance of the meta learner. LOOCV is particularly suitable for small datasets, as it maximizes the use of available training data by designating a single sample as the validation set and using the remaining samples for training [31]. This can be regarded as a special case of cross-validation in which the validation set consists of only one observation.

Figure 7 illustrates the validation set partitioning strategies employed in this study to evaluate the performance of support vector machines. For the nine SVMs used as base learners, two types of validation approaches are adopted. As shown in Figure 7(a), four SVMs and all neural networks in the base learner use a general partitioning validation strategy. In contrast, Figure 7(b) illustrates the use of nested cross-validation for the remaining five SVMs, where only the outer cross-validation loop is depicted for clarity. Finally, the meta learner utilizes LOOCV, as shown in Figure 7(c), ensuring a robust and unbiased evaluation based on minimal data partitioning.

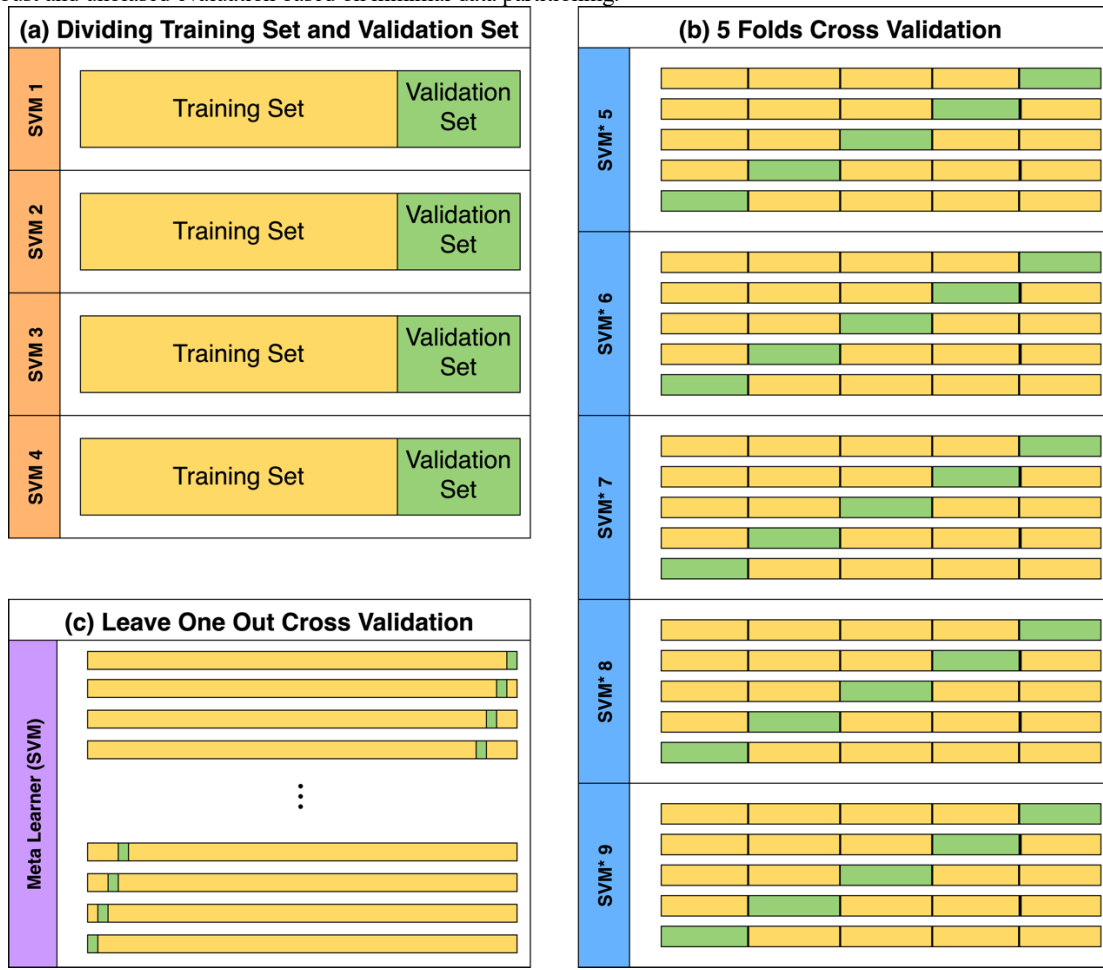

Figure 6. Leave-one-out cross validation set arrangement.

### 4.4 Early stopping

The early stopping strategy refers to the practice of halting neural network training before the model fully converges on the training set, typically prior to reaching its optimal performance on the training data [32]. As a widely recognized and effective technique for preventing overfitting, early stopping is also considered a form of regularization [33]. The concept was first introduced into machine learning by H. Morgan et al. in 1990 [34].

The core idea of early stopping is to prevent the neural network from overfitting the training data by discouraging it from memorizing specific features of the samples. Instead, it encourages the model to learn more generalizable patterns that better reflect the underlying data distribution. In this study, early stopping is employed during the training of the proposed model to mitigate overfitting, with particular attention paid to determining the optimal stopping point for maximum generalization performance.

## 5 EXPERIMENTS WITH PUBLIC DATA SETS

The experimental workflow adopted in this study is illustrated in Figure 8, which shows that prior to reaching the ensemble learning model, the data undergoes a series of sequential processing steps—comparable to an assembly line—which ultimately produces the final emotion recognition results. The following section provides a detailed explanation of the key stages outlined in the flowchart, along with additional procedures not explicitly depicted in the figure.

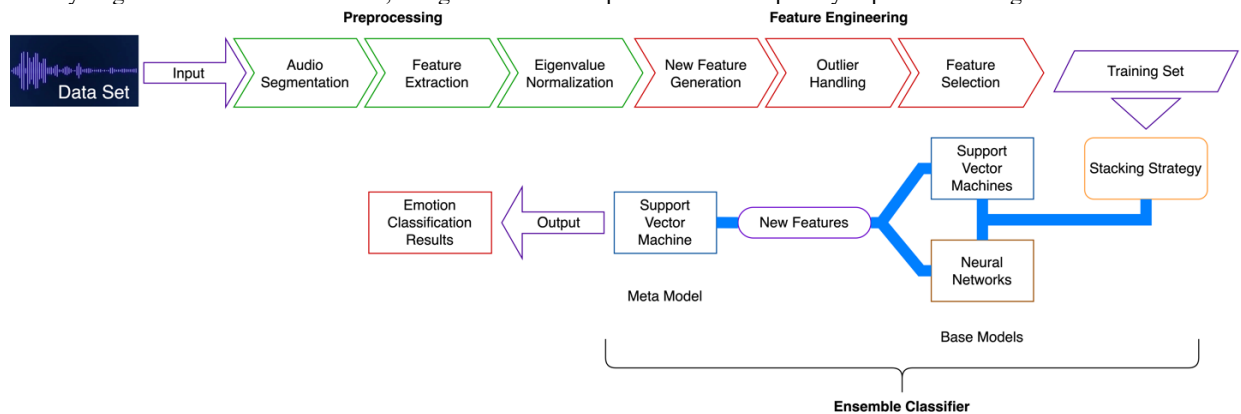

Figure 7．Experimental flowchart of this study.

### 5.1 Preparing the public data sets

We hypothesize that if the proposed ensemble learning model can accurately classify emotions within the publicly available datasets, it would then be worthwhile to invest resources in recruiting annotators for the development of a privately prepared movie audio emotion dataset.

The Turkish Music Emotion Dataset (TMED) dataset [35] comprises verbal and nonverbal emotional expressions across various genres of Turkish music. It includes four emotion categories—happy, sad, angry, and relaxed—with 100 equally distributed samples per class. The Audio Emotions Dataset is a composite corpus that integrates four prominent speech emotion datasets: RAVDESS (58.15%) [36], CREMA-D (21.88%) [37], SAVEE (16.22%) [38], and TESS (3.75%) [39]. It covers seven emotion categories: angry, happy, sad, neutral, fearful, disgusted, and surprised. For this study, we aim to construct a simulated movie audio emotion dataset with three overarching emotion classes—good, neutral, and bad—by reorganizing and relabeling the categories from the aforementioned publicly available datasets.

Table 3．Combination of emotion categories for public data sets

| Emotion | Category of Turkish Music Emotion Dataset | Category of Audio emotions |
|---|---|---|
| Good | Happy | Relax |
|  | Happy | Surprised |
| Neutral | None | Neutral |
| Bad | Angry | Sad |
|  | Angry | Fearful |

Table 3 shows the combining strategies for emotion categories in the two available datasets. The simulated dataset generates good emotion recordings by combining TMED's Happy label samples with Audio Emotions' Relax and Surprised label samples. Similarly, Bad emotion samples are made by merging TMED's Angry label samples with Audio Emotions' Sad and Fearful labels. Because there is no equivalent emotion music in TMED, the Neutral emotion samples from Audio Emotions are used directly in the simulation dataset. In the studies, the overlay() method from Python's Audio Segment library was used to combine samples from both datasets into a single WAV audio file, labeling the fused samples as Good, Neutral, and Bad. Thirty samples were selected at random from each group, yielding 90 samples for each of the three categories and constituting a simulated dataset.

Table 3 presents the emotion category mapping strategies applied to the two publicly available datasets. To simulate Good emotion samples, recordings labeled as Happy from the TMED dataset were combined with Relaxed and Surprised samples from the Audio Emotions dataset. Bad emotion samples were created by merging Angry samples from TMED with Sad and Fearful samples from Audio Emotions. As the TMED dataset lacks a direct counterpart for neutral emotional content, Neutral samples from the Audio Emotions dataset were used directly to represent the Neutral category in the simulated dataset.

To generate the final audio samples, the overlay() method from Python's AudioSegment library was employed to blend samples from the speech and music datasets into single WAV files, each labeled as Good, Neutral, or Bad. A total of 30 samples were randomly selected for each emotion class, resulting in 90 recordings that collectively form the simulated movie audio emotion dataset.

## 5.2 Preprocessing of data

*5.2.1 Segmentation and feature extraction of audio signal*

In real-time signal processing tasks—particularly when dealing with continuous temporal signals such as audio—windowing operations are essential for defining the smallest unit of analysis. Following precedents established in earlier studies, this investigation adopts a 7-second window duration as the basic processing unit [2]. Accordingly, all audio samples are segmented into 7-second clips, with each segment serving as the fundamental unit for emotion classification.

To extract emotional information from these 7-second segments, a set of audio features must be computed. These features can be broadly categorized into three domains: Cepstral Domain, Time Domain, and Frequency Domain. Table 4 presents representative features from each category. Among these, the Mel-Frequency Cepstral Coefficients (MFCCs) from the Cepstral Domain are particularly important. MFCCs are widely used and highly effective features for one-dimensional audio signals [40, 41]. Their design is inspired by the human auditory system, which exhibits limited sensitivity to frequencies above 1 kHz [42]. Consequently, MFCCs focus on capturing information within the frequency range that aligns with human perceptual capabilities [43], making them especially well-suited for tasks involving speech and emotion recognition.

Table 4. The features selected on the three domains.

| Cepstral Domain | Time Domain | Frequency Domain |
|---|---|---|
| MFCC 0 to 23 | Zero Crossing Rate | Spectral Centroid |
| Melspectrogram | | Chroma STFT 0 to 11 |
| | | Spectral Roll-off |

The calculation of the MFCC consists of two steps such as mel-frequency wrapping and cepstrum converting.

1. Mel-frequency wrapping

Accurately simulating the human auditory system is critical for effective audio feature extraction. Unlike linear frequency scales, human perception of sound operates on a non-linear subjective scale known as the Mel scale. This scale transforms actual frequencies into perceptual pitch representations, more closely aligning with how humans interpret variations in sound. The Mel-frequency scale was developed to better reflect this subjective perception by adjusting the spacing between adjacent frequencies according to human auditory sensitivity. This process—known as Mel-frequency wrapping—converts the frequency axis of an audio signal into the Mel scale. Given a frequency $f$ (Hz), the perceived frequency in Mels can be approximated by,

$$\mathrm{Mel}(f) = 2595 * \log 10(1 + f/700) \tag{2}$$

Mel(f) captures the nonlinear relationship between actual frequency and perceived pitch, aligning closely with the characteristics of human auditory perception. To better reflect the subjective auditory spectrum, this study employs a Mel-scale filter bank consisting of 24 filters to transform the audio signal into a set of Mel-scaled frequency components. Each audio sample is thus represented by 24 Mel coefficients, providing a richer and more perceptually accurate representation of the audio content.

2. Cepstrum inversion

The previously obtained log Mel spectrum is converted back to the time domain in order to produce the Mel Frequency Cepstrum Coefficients (MFCC). The discrete cosine transformation, as described in Eq. (3), is used to convert from the cepstral domain to the time domain.

$$C_n = \sum_{k=1}^{k} (\log S_k) \cos n \left(k - \frac{1}{2}\right) * \frac{\pi}{k}, n_1 = 1,2,\ldots k \tag{3}$$

where $S_k, K = 1,2,\ldots K$ is the output of the previous step.

The Melspectrogram is obtained by filtering the results of the short-time Fourier transform of the audio signal with a bank of Mel-frequency filters [44]. The coefficients $S_{mel}(k, t)$ of the Melspectrogram are computed by Eq. (4):

$$S_{mel}(k, t) = \sum_{l=0}^{L-1} m_k(l) |S(l, t)^2| \tag{4}$$

where L is the frequency component number. $m_k(l)$ denotes the kth filter that responds on component l.

Zero Crossing Rate is the frequency at which an audio signal crosses a value of 0 for a given period of time [45]. Its value is calculated by Eq. (5), such that

$$Z(i) = \frac{1}{2W_L} \sum_{n=1}^{W_L} | \text{sgn}[x_i(n)] - \text{sgn}[x_i(n-1)] | \tag{5}$$

Where the sgn is $\text{sgn}[x_i(n)] = \begin{cases} 1, & x_i(n) \geq 0, \\ -1, & x_i(n) < 0. \end{cases}$ . This formula describes how to calculate the Zero Crossing Rate $Z(i)$ for an audio signal $x_i(n)$. i is the ith sub-segment of the signal, $W_L$ is the length of the sub-segment, and n denotes the n th sampling point [45].

Spectral centroid. The spectral center of mass is a measure of where the spectrum's center is located, represented by the average value of the weights of each frequency component. To find the spectral center of mass, multiply the frequency value of each frequency component by its amplitude value in the spectrum. This means that for each frequency in the spectrum, the product of its frequency and magnitude values is computed. The products of all frequency components are then combined together to form a total. This is the weighted sum of all frequency components in the amplitude spectrum. This calculation process can be described by Eq. (6).

$$Sc = \frac{\left(\sum_{i=1}^{N} f(i) \times M[f(i)]\right)}{\sum_{i=1}^{N} M[f(i)]} \tag{6}$$

where represents the amplitude at frequency in the th frequency band. represents the number of frequency bands in the spectrum, or frequency resolution [46–48].

Humans perceive pitches to be similar in a cyclical sense, which means that if two pitches differ by one or more octaves, they are considered to perform the same harmonic role. The concept of color is utilized here to convey pitch similarity rather than visual color. Pitch consists of two components: tone height and chroma. In other words, chroma is a technique for characterizing pitch [49] and was chosen as one of the frequency domain variables in this investigation. After using 12 Chroma filters, a total of 12 Chroma characteristics were collected.

Spectral roll-off is a measure that describes the energy-frequency connection in a spectrum. It represents the proportion of total energy below a specific frequency in the spectrum, commonly given as a percentage. In this situation, spectral attenuation can be used to calculate the proportion of energy in the high-frequency component of the spoken signal in relation to overall energy. If the spectral attenuation percentage is large, the energy in the high-frequency region is low, and vice versa [50]. Given that the movie contains a portion of the speech signal, and that the energy in the speech signal is frequently distributed in the low-frequency range, with energy gradually decreasing as frequency increases [51], we chose spectral roll-off as a feature to characterize the relationship between energy and frequency.

*5.2.2 Outlier handling*

Hawkins defines an outlier as an observation that significantly deviates from the rest of the data, arising from a distinct mechanism that differs from the one generating the normal values [52]. Machine learning algorithms aim to fit an optimal model while avoiding the incorporation of erroneous information from outliers. Additionally, outliers can diminish the accuracy of classification algorithms and negatively impact data analysis [53]. Therefore, it is crucial to address outliers prior to progressing to the next stage of data processing.

The Boxplot method, introduced by John Tukey, is commonly used to identify both minor and extreme outliers [54]. However, in this study, the degree of deviation of the outliers is not of primary concern. Instead, the Boxplot is employed to determine whether a given eigenvalue falls within the interquartile range (IQR), which represents the distribution of normal values. If an eigenvalue lies outside this range, it is flagged as an outlier and replaced by the median value of the corresponding feature [55].

Figure 9 illustrates an example of a box-and-whisker plot of features. The plot shows that at least five feature values are extreme outliers, as their positions lie outside the interquartile range (represented by the vertical lines) near the x-axis values of -50 and 10, respectively. In this visualization, the blue box encapsulates 50% of the eigenvalues, while values outside the box but within the IQR are considered minor outliers. In this particular feature, no minor outliers are present.

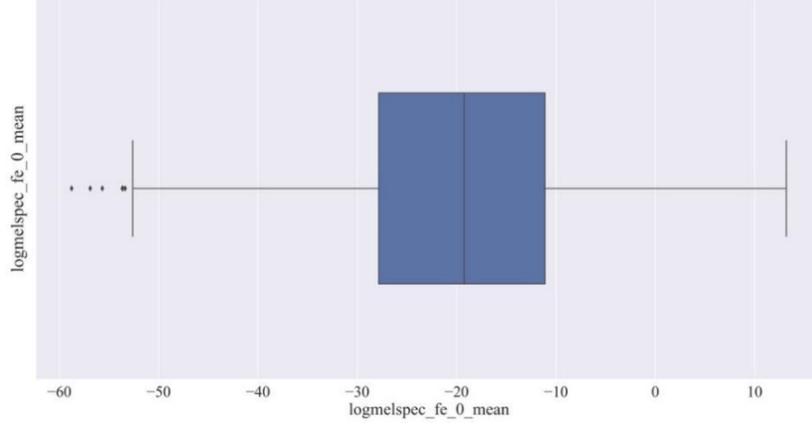

Figure 8. Boxplot of an example feature.

*5.2.3 Eigenvalue normalization*

Data normalization is a crucial step in the data preprocessing pipeline, which involves adjusting the values of attributes to a consistent scale or range. This process enhances the performance of machine learning algorithms [56]. Failure to perform feature scaling may lead to certain values being overlooked, potentially compromising the efficiency and accuracy of the algorithm [57, 58]. Therefore, Eq.(7) is used to normalize the features through the Min-Max Scaling method, as follows:

$$X_{scaled} = \frac{X - X_{min}}{X_{max} - X_{min}} \quad (7)$$

After normalization, the values of all features are bounded in the interval [0, 1].

*5.2.4 Data set partitioning*

In machine learning workflows, datasets are typically divided into three subsets: training, validation, and testing. The training set is used to train the classification algorithm, while the validation set is employed to monitor the system's performance during training and fine-tune the model's parameters. The test set is reserved for the final evaluation of the model's performance [59]. In this study, the validation set is not initially separated; instead, a portion of it is isolated from the training set during specific training steps. The test set, however, is partitioned before training begins and remains unseen by the model throughout the training process, serving exclusively for the final evaluation of the model's performance. The test set comprises 15% of the samples, randomly selected from the entire dataset. The proportion of the validation set varies depending on the validation strategy employed during the learner's training, which will be discussed in the subsequent training steps.

**5.3 Feature engineering**

*5.3.1 New feature generation*

The learner's prediction performance can be improved by new characteristics acquired through suitable construction techniques [60]. Three techniques for creating new features from existing features are presented in this study such as mean feature, range (diff) and mean absolute deviation (MAD).

Mean feature gives an indication of the concentration trend of a physical feature value in a sample by calculating the average of several values of that feature [61]; Range indicates the degree of dispersion and can be calculated as,

$$\text{Range} = x_{max} - x_{min} \tag{8}$$

While a smaller value of Range implies a more concentrated distribution, a larger value denotes a more discontinuous distribution of the characteristic [62]; MAD evaluates how much each variable's values vary from one another, such that,

$$\text{MAD} = \frac{1}{n}\sum_{i=1}^{n} |x_i - \text{mean}(X)| \tag{9}$$

After the feature generation process described above, the final dataset contains 195 features which include features identified as outliers. The following outlier management procedure will address these outliers to ensure their proper handling.

*5.3.2 Feature selection*

An excessive number of features increases the risk of overfitting, elevates computational costs, and may lead to the "curse of dimensionality." Therefore, feature selection is especially critical for high-dimensional datasets to reduce dimensionality effectively [63]. Proper feature selection enables the learner to achieve strong performance using fewer features, thereby reducing computational effort and lowering hardware requirements [64, 65]. Accordingly, Figure 10 illustrates the set of filter pipelines designed for this study. This filter pool comprises four sequential filters, each progressively reducing the dimensionality of the input features. The following sections describe the function and principles of each filter in detail.

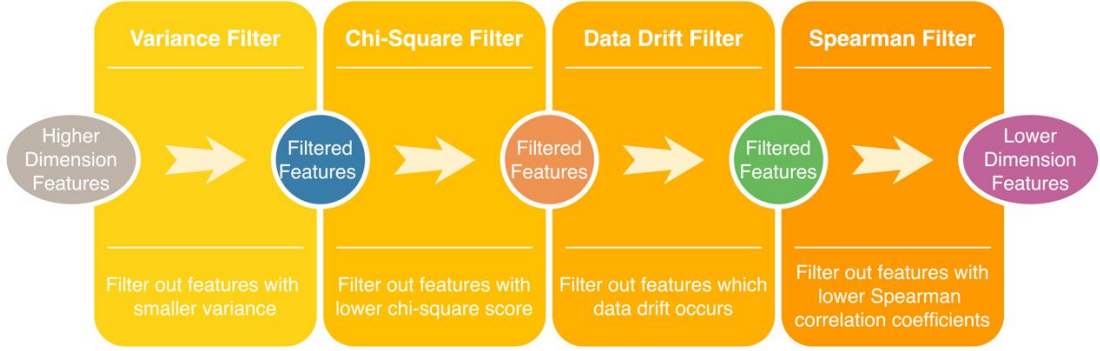

Figure 9. Filters bank

The first filter in the pipeline is the Variance Filter, which is designed to eliminate features with variance below a specified threshold value. Variance assesses the degree of dispersion within a dataset such that,

$$\sigma^2 = \frac{\sum(x-\mu)^2}{N} \tag{10}$$

where μ is the mean value and N represents the sample size [66]. In general, features with larger variance are usually considered to better reflect the variability among different samples. The threshold can be set to 0.02 based on experiment.

The Chi-Square Filter is a second filter that aims to eliminate features with small chi-square test scores. The chi-square test is a statistical method used to assess the degree to which features deviate from category independence and can be computed such that [67, 68],

$$\chi^2 = \sum_{i=1}^{n} \frac{(E_i - O_i)^2}{E_i} \tag{11}$$

where $E_i$ represents the expected value of the occurrence of the i th feature within a specific category, and $O_i$ denotes the actual occurrence of the i th sample in that particular category. This filter computes the Chi-square test scores for each feature and subsequently arranges them in descending order, retaining only the top 60 features with higher chi-square test scores.

This nonparametric estimation method avoids making assumptions about the underlying densities, allowing the data to manifest its own distribution [74]. The kernel density estimator can be given as,

$$\widehat{f}_h(x) = \frac{1}{nh}\sum_{i=1}^{n} K\left(\frac{x-x_i}{h}\right) \tag{12}$$

Where K is a kernel function and h is a smoothing parameter (bandwidth). Typically, the kernel function used in Kernel Density Estimation (KDE) is a Gaussian function centered at zero, resulting in a single-peaked, symmetric estimator about this point [74]. Accordingly, this study employs KDE with a Gaussian kernel to visualize the distribution of each feature. Figure 11 displays the KDE plots for six selected features, illustrating how these visualizations enable comparison of sample distributions between the training and test sets and facilitate the detection of potential data drift.

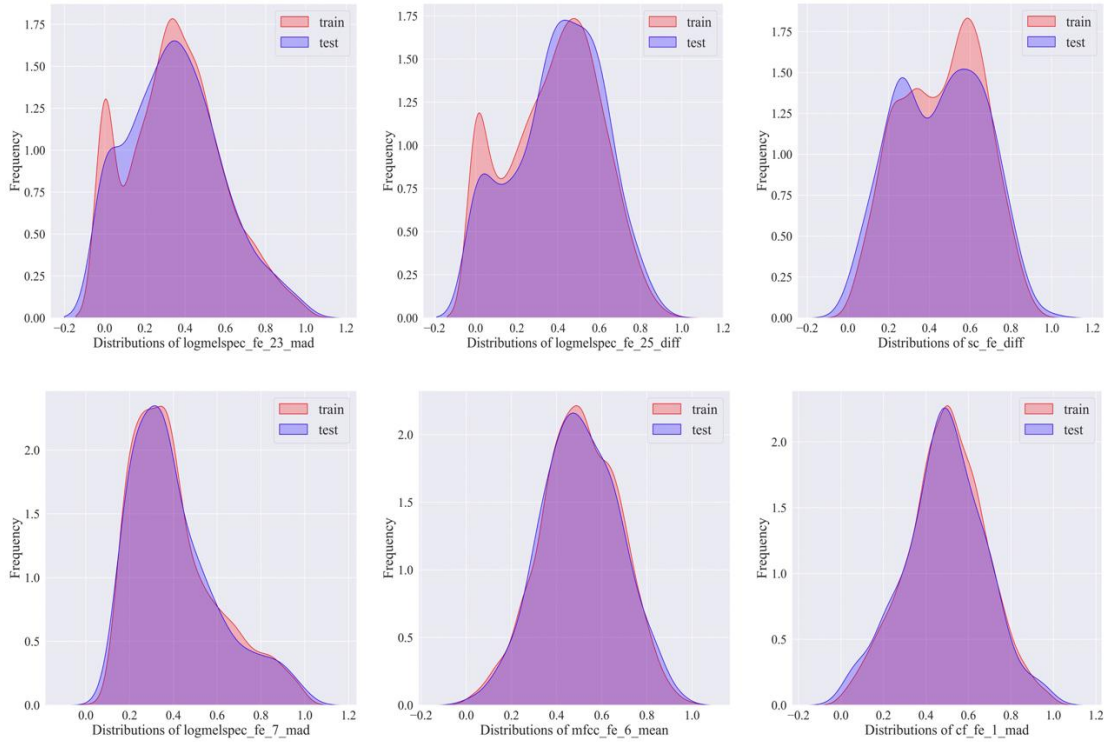

Figure 10．Examples of KDE plots.

The three subplots in the first row of Figure 11 illustrate notable differences in feature distributions between the training and test datasets across the sample space. Since many machine learning models assume that input samples are independent and identically distributed (i.i.d.), such discrepancies indicate the presence of data drift from the training to the test set, which may adversely affect model performance. In contrast, the second row of Figure 11 displays features whose distributions exhibit substantial similarity between the training and test sets, as evidenced by the significant overlap between the blue and red regions. This consistency aligns with the expectation that features should be similarly distributed across both datasets. Based on this approach, KDE plots will be generated for all features. After manual review, features showing considerable overlap between training and test distributions will be retained, while those with minimal overlap will be excluded from further analysis.

The Spearman filter serves as the final stage in feature selection by eliminating features whose correlation with the target labels falls below a specified threshold. Not all features extracted from audio samples exhibit a meaningful relationship with the labels. Features that are irrelevant or show low correlation with the categories may provide little value to the classification algorithm and can potentially degrade its performance [75]. To address this, the Spearman correlation coefficient is employed to quantify the strength of association between each feature and the labels. Specifically, this filter calculates the Spearman correlation coefficient for each feature-label pair, defined as follows [76]:

$$r_s = \frac{\sum_{i=1}^{N} x_{i,r} y_{i,r}}{\sqrt{\sum_{i=1}^{N} x_{i,r}^2 \sum_{i=1}^{N} y_{i,r}^2}} \tag{13}$$

Where the subscript $r_s$ denotes the ranked value. The threshold is set at 0.08 to satisfy the filtering criterion. Figure 12 shows the correlation heatmap between features and emotions of the filtering output.

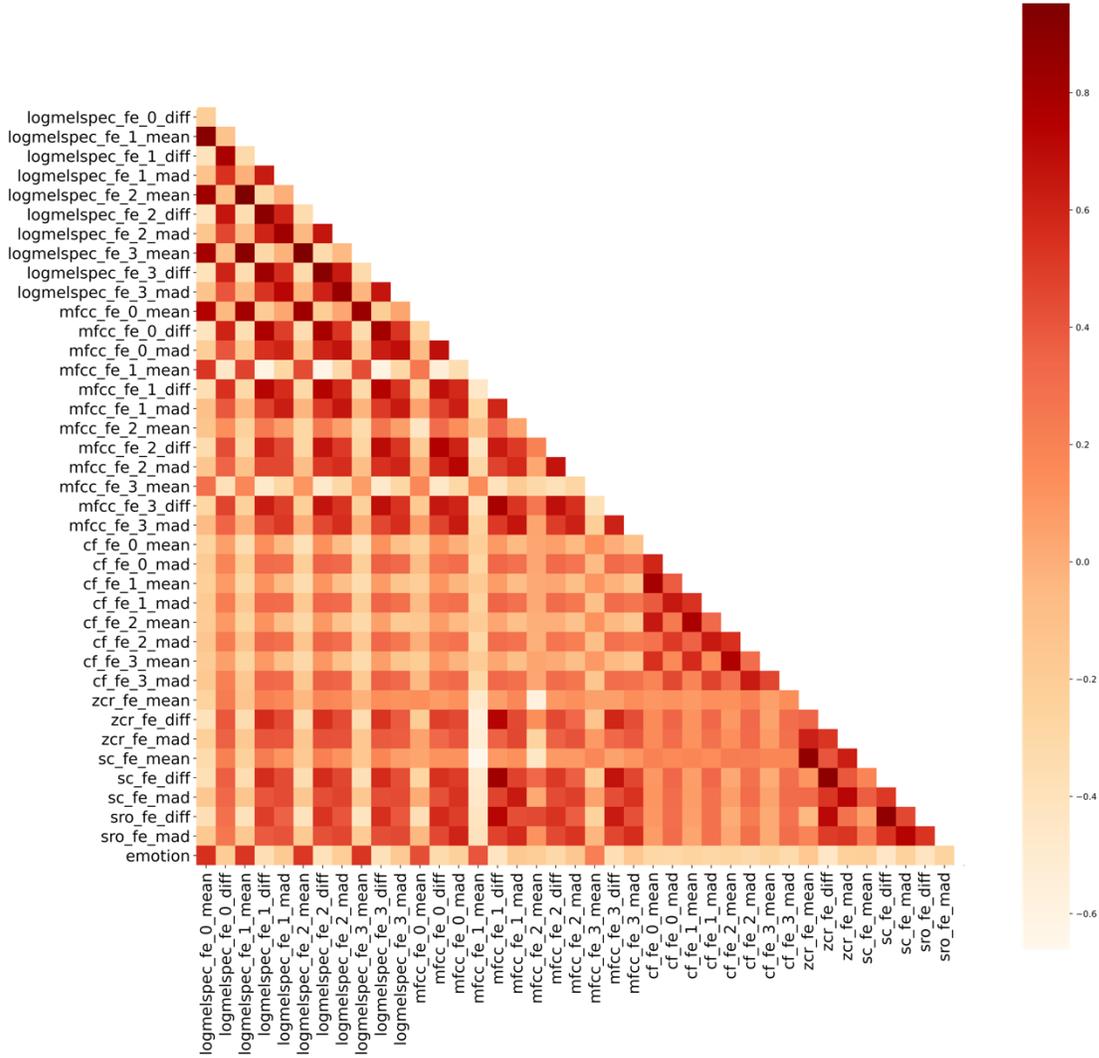

Figure 11. Correlation heat map between features and emotions.

## 5.4 Model training and its evaluation results

The proposed ensemble learning model is trained and tested on the public datasets, and the test results for the three types of emotion categories are evaluated with Precision, Recall and F1 respectively, such that,

$$\text{Precision} = \frac{TP}{TP + FP} \tag{14}$$

$$\text{Recall} = \frac{TP}{TP + FN} \tag{15}$$

$$F1 = 2 \cdot \frac{\text{Precision} \cdot \text{Recall}}{\text{Precison} + \text{Recall}} \tag{16}$$

Where TP is true positive class samples that were accurately projected as positive; FP denotes false positive class samples that were incorrectly forecasted as positive; FN represents false negative class samples that were wrongly predicted as negative. In a three-classification problem, one class is classified as positive, and the remaining classes are regarded as negative.

Table 5．Emotion recognition results of ensemble learning models on public data sets.

| Emotion | Results on Test Set | | |
| --- | --- | --- | --- |
| | Precision | Recall | F1 |
| G | 0.73 | 0.63 | 0.68 |
| N | 0.62 | 0.70 | 0.65 |
| B | 0.67 | 0.66 | 0.66 |
| Accuracy: 0.67 | | | |

The overall recognition accuracy across all emotion categories is 67%. According to classical probability theory, random guessing in a three-class classification problem would yield an accuracy of approximately 33%. The observed accuracy, being twice that of random chance, indicates that our ensemble learning approach has significant potential to deliver meaningful results on real-world movie audio emotion datasets. Experiments with private dataset

**5.5 Preparation of private data sets**

*5.5.1 Movie Audio Extraction*

To further evaluate the generalization capability of the ensemble learning model, a private dataset was meticulously compiled and applied to the proposed model. This dataset consists of audio extracted from fifteen films spanning a diverse array of languages, including English, Mandarin, Cantonese, Korean, Italian, Japanese, German, and Swedish. The selected films encompass a broad spectrum of genres such as drama, narrative, crime, comedy, suspense, thriller, and fantasy. Detailed information about these movies is provided in Table 6. Following the methodology employed for the simulated dataset, the audio signals extracted from these films were segmented into samples of uniform length, each with a 7-second window duration.

Table 6. The film source of the private dataset.

| Title | Genres | Language | Country and/or Region | Release Date |
| --- | --- | --- | --- | --- |
| Farewell My Concubine | Drama | Mandarin | China Mainland & H.K. | Jul. 26, 1993 |
| A Man Called Ove | Narrative | Swedish | Sweden | Dec. 25, 2015 |
| The God Father | Crime | English | U.S.A. | Mar. 15, 1972 |
| Life is Beautiful | Narrative | Italian | Italy | Dec. 20, 1997 |
| King of Comedy | Comedy | Cantonese | Hong Kong | Feb. 13, 1999 |
| So-won | Suspense | Korean | South Korea | Oct. 2, 2013 |
| Lord of War | Crime | English | U.S.A. & Germany | Sep. 16, 2005 |
| To Live | Narrative | Mandarin | China Mainland & H.K. | May. 17, 1994 |
| Identity | Thriller | English | U.S.A. | Apr. 25, 2003 |
| Spirited Away | Fantasy | Japanese | Japan | Jul. 20, 2001 |
| Comrades: Almost a Love Story | Narrative | Cantonese | Hong Kong | Nov. 2, 1996 |
| Forrest Gump | Narrative | English | U.S.A. | Jun. 23, 1994 |
| Eat Drink Man Woman | Narrative | Mandarin | Taiwan & U.S.A. | Jul. 2, 1994 |
| The Invisible Guest | Thriller | Spanish | Spain | Sep. 23, 2016 |
| The Lives of Others | Suspense | German | Germany | Mar. 23, 2006 |

*5.5.2 Emotion Labeling*

Each sample in the private dataset was labeled as one of three emotions: positive, neutral, or negative. This labeling task was carried out by three volunteers, each holding at least a bachelor's degree. To facilitate the annotation process, a standardized form (see Figure 18 in the Appendix) was developed. This form captures essential details about the task and records the labeling outcomes for each annotation cycle. Each sample was independently labeled by all three volunteers, and the final emotion label was determined using a majority voting rule. Samples for which all three annotators provided differing labels were deemed inconclusive and subsequently excluded from the dataset.

**Table 7.** Details of the labeling volunteers.

| ID | Gender | Age | Education Level | Cultural Background | Native Language | Foreign Language (proficiency) |
|---|---|---|---|---|---|---|
| 0 | Female | 21 | Undergraduate | Cantonese | Cantonese & Mandarin | English (Command) |
| 1 | Female | 22 | Postgraduate | Mandarin | Mandarin | English (Master) & Cantonese (Command) |
| 2 | Female | 20 | Undergraduate | English | English & Russian | None |

Volunteer 0 primarily speaks Cantonese in daily life, except when teaching Mandarin in schools, and has basic proficiency in English across speaking, listening, reading, and writing. Volunteer 1 is a native Mandarin speaker fluent in both Mandarin and English in all contexts, including at home, with a strong background in journalism and high proficiency in both languages. Volunteer 2 is a native English speaker with additional proficiency in Russian.

The rationale for selecting labeling volunteers from diverse linguistic backgrounds stems from the need to mitigate the influence of actors' speech content embedded in the audio streams on the labeling outcomes. For example, in the Cantonese version of King of Comedy, which features humorous actor dialogues, Volunteer 0—fluent in Cantonese—may assign a "Good" emotion label reflecting the comedic tone. In contrast, Volunteer 2, lacking Cantonese comprehension, might perceive the same exchanges as neutral or mundane, thereby assigning a "Neutral" label. Similarly, for German-language films such as The Lives of Others, where none of the volunteers are fluent in German, emotion labels are based primarily on auditory cues and intuition. For instance, a German dialogue paired with a disturbing soundtrack may be labeled "Bad" due to the unsettling music, irrespective of the dialogue's semantic content. Likewise, in Comrades: Almost a Love Story, scenes featuring soothing background music alongside the protagonists' melancholic conversations can evoke divergent interpretations: Volunteer 1 and Volunteer 2, with limited Cantonese understanding, might rate these scenes positively, while Volunteer 0, who comprehends the dialogue, may interpret them negatively. The inclusion of volunteers with varied linguistic and cultural backgrounds enhances the robustness and generalizability of the labeling process. Using this approach, a total of 9 hours, 54 minutes, and 24 seconds of movie audio has been annotated with emotion labels.

*5.5.3Balancing the data categories*

Following the labeling process, a significant imbalance was observed in the distribution of emotion categories across the samples, with a large proportion classified as neutral. Such class imbalance can lead classification algorithms to bias learning toward the majority class, thereby neglecting the minority classes and their distinguishing features [77]. To mitigate this issue, it is essential to balance the sample counts across the Good, Neutral, and Bad categories.

In this study, the Near Miss algorithm is employed to achieve dataset balancing. Near Miss is an undersampling technique that considers the spatial distribution of samples among multiple classes. Specifically, when sample points from both majority and minority classes are densely clustered, the algorithm selectively removes samples from the majority class [78]. This undersampling process is performed iteratively until a numerical balance is established among the three emotion categories.

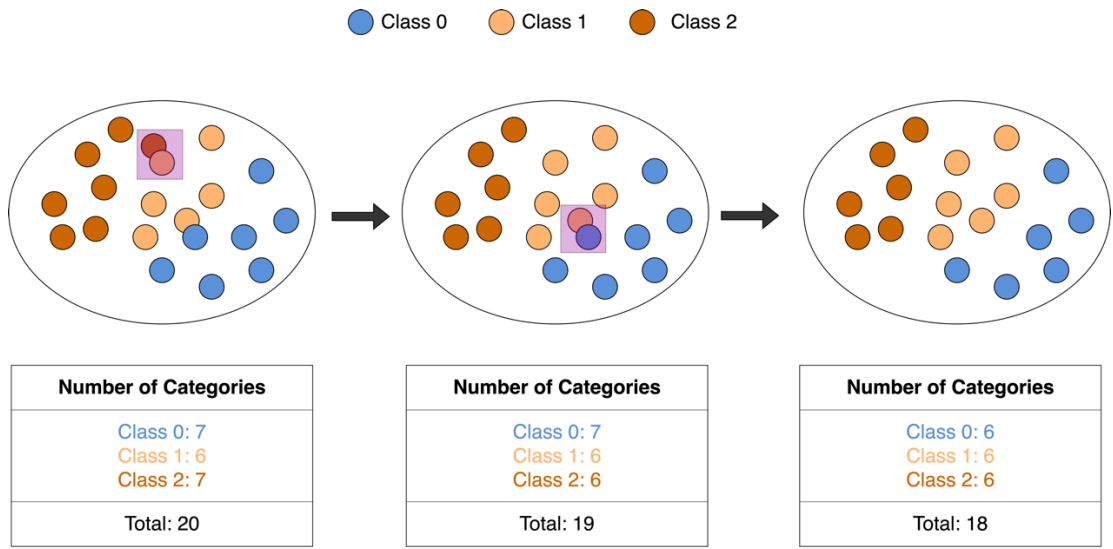

Figure 12. The Near Miss algorithm.

Figure 13 illustrates a schematic of the undersampling process applied to majority classes located near minority class samples. In an imbalanced dataset, Class 1 represents the minority class with fewer samples compared to Classes 0 and 2, which constitute the majority classes. The algorithm identifies samples from Classes 0 and 2 that are closest to the distribution of Class 1 and removes them. In the figure's legend, majority class samples nearest to the minority class are highlighted within a translucent purple box, indicating the subset targeted for elimination by the algorithm. By employing this nearest-neighbor minority class undersampling technique, the sample ratio across the three emotion classes—Good, Neutral, and Bad—is balanced to 1:1:1. Such an even distribution promotes more effective learning by the classification algorithm.

## 5.6 The training of learners

*5.6.1 The training of base SVMs*

The hyperparameter optimization process begins with a K-fold cross-validation grid search. As detailed in Round 1 of Table 8, the initial search space for the penalty parameter C and kernel coefficient γ is defined within the closed interval of 0.1 to 10. Recognizing that higher values of γ and C can increase the risk of overfitting in support vector machines, a denser distribution of candidate values is selected towards the lower end of this interval. In the first search round, the grid search identified a combination of C and γ that yielded an 86% cross-validation accuracy.

Building on this result, a second, more focused search was conducted in the neighborhood of the previously identified optimal parameters. This refinement led to significant improvements in the recall and F1 score for the neutral emotion class, which rose to 90% and 0.88, respectively, while the overall cross-validation accuracy remained stable. Furthermore, the accuracy and F1 score for the Bad emotion class showed notable gains, increasing to 88% and 0.87, respectively. Although there was a slight decline in the recall and F1 score for the Good emotion class, the overall enhancements in classification performance are considered satisfactory.

Table 8. Grid search process 1.

| Round | Scope of C | Scope of γ | Better Parameters | Emo | P | R | F1 |
|---|---|---|---|---|---|---|---|
| 1 | [0.1, 0.5, 1, 2, 3, 5, 7, 10] | [0.1, 0.5, 1, 2, 3, 5, 7, 10] | C = 2, γ = 0.5 | G | 0.85 | 0.85 | 0.85 |
| | | | | N | 0.87 | 0.88 | 0.87 |
| | | | | B | 0.86 | 0.86 | 0.86 |
| | | | | Accuracy: 0.86 | | | |
| 2 | [1.5, 2, 2.5] | [0.3, 0.5, 0.7] | C = 2.5, γ = 0.5 | G | 0.85 | 0.84 | 0.84 |
| | | | | N | 0.87 | 0.90 | 0.88 |
| | | | | B | 0.88 | 0.86 | 0.87 |
| | | | | Accuracy: 0.86 | | | |
| 3 | [2.2, 2.35, 2.5, 2.65, 2.8] | [0.35, 0.5, 0.65] | C = 2.35, γ = 0.5 | G | 0.85 | 0.84 | 0.84 |
| | | | | N | 0.87 | 0.90 | 0.88 |
| | | | | B | 0.88 | 0.86 | 0.87 |
| | | | | Accuracy: 0.86 | | | |
| 4 | [2.3, 2.4] | [0.5] | C = 2.3, γ = 0.5 | G | 0.85 | 0.84 | 0.84 |
| | | | | N | 0.87 | 0.90 | 0.88 |
| | | | | B | 0.88 | 0.86 | 0.87 |
| | | | | Accuracy: 0.86 | | | |

Table 8 provides a detailed account of the complete hyperparameter search process. Despite conducting iterative searches during the third, fourth, and fifth rounds, the performance of the SVMs trained with the updated hyperparameter combinations showed no significant variation. Consequently, the final optimized values were set to C=2.3C = 2.3C=2.3 and γ=0.5\gamma = 0.5γ=0.5. The inclusion of four K-fold cross-validated SVMs in the base learner corresponds to the construction of four distinct SVM models obtained across the four search rounds.

Subsequently, these SVMs undergo further hyperparameter optimization via a grid search embedded within a nested cross-validation framework. The initial hyperparameter search space again lies within the closed interval of 0.1 to 10. The model's performance is evaluated using the lattice search algorithm in conjunction with nested cross-validation, with the results summarized in Table 9, which details the search procedure.

As shown in Table 9, the hyperparameters C and γ selected in the fifth round were 3 and 0.6, respectively, yielding a cross-validation accuracy of 85.3%. Although this accuracy is slightly lower than the 86% achieved in the previous search cycle (Table 8), the results from nested cross-validation are generally considered more reliable and better representative of real-world performance. Based on the hyperparameters obtained from the five rounds of grid search within the nested cross-validation framework, five SVM models were constructed to serve as base learners in the ensemble.

Table 9. Grid search process 2.

| Round | Scope of C | Scope of γ | Better Parameters | Results on Validation Set | | | |
|---|---|---|---|---|---|---|---|
| | | | | Emo | P | R | F1 |
| 1* | [0.1, 0.5, 1, 2, 3, 5, 7, 10] | [0.1, 0.5, 1, 2, 3, 5, 7, 10] | C = 2 γ = 0.5 | G | 0.85 | 0.85 | 0.85 |
| | | | | N | 0.87 | 0.88 | 0.87 |
| | | | | B | 0.86 | 0.86 | 0.86 |
| | | | | Accuracy: 0.851 | | | |
| 2* | [1.5, 2, 2.5] | [0.3, 0.5, 0.7] | C = 2.5 γ = 0.5 | G | 0.86 | 0.84 | 0.85 |
| | | | | N | 0.88 | 0.88 | 0.88 |
| | | | | B | 0.84 | 0.87 | 0.85 |
| | | | | Accuracy: 0.85 | | | |
| 3* | [1.5, 2, 2.5] | [0.6, 0.7, 0.8] | C = 2.5 γ = 0.6 | G | 0.86 | 0.84 | 0.85 |
| | | | | N | 0.87 | 0.88 | 0.87 |
| | | | | B | 0.85 | 0.86 | 0.85 |
| | | | | Accuracy: 0.85 | | | |
| 4* | [2.5, 3] | [0.5, 0.6] | C = 3 γ = 0.6 | G | 0.86 | 0.84 | 0.85 |
| | | | | N | 0.87 | 0.88 | 0.87 |
| | | | | B | 0.85 | 0.86 | 0.86 |
| | | | | Accuracy: 0.853 | | | |
| 5* | [3, 3.5] | [0.6] | C = 3 γ = 0.6 | G | 0.86 | 0.84 | 0.85 |
| | | | | N | 0.87 | 0.88 | 0.87 |
| | | | | B | 0.85 | 0.86 | 0.86 |
| | | | | Accuracy: 0.853 | | | |

The hyperparameters selected in the final search round represent approximations near the optimal solution. As detailed in Table 10, the support vector machine configured with these parameters was chosen for performance evaluation on the test set. Notably, the two SVM models exhibited alarmingly low classification accuracy—below 50%—for the Bad emotion class, which is a cause for concern. Although the overall accuracy marginally exceeds 70%, the classification performance for the Good and Neutral classes remains satisfactory. These results suggest that, despite utilizing near-optimal hyperparameter combinations identified in the final search, the SVMs still demonstrate inherent limitations as classifiers, particularly in distinguishing the Bad emotion category.

Table 10. The performance of SVMs on test set.

| Grid Search Process | Emotion | Precision | Recall | F1 | Accuracy |
|---|---|---|---|---|---|
| 1 | G | 0.81 | 0.72 | 0.76 | |
| | N | 0.94 | 0.67 | 0.79 | 0.73 |
| | B | 0.48 | 0.87 | 0.62 | |
|   | G | 0.83 | 0.70 | 0.76 |   |
| 2 | N | 0.94 | 0.65 | 0.77 | 0.71 |
|   | B | 0.46 | 0.88 | 0.61 |   |

*5.6.2 Training of base neural networks*

This section details the training of the Backpropagation Neural Network (BPNN) and one-dimensional Convolutional Neural Networks (1D-CNNs) within the base learner. The validation set, depicted in the upper-left subfigure of Figure 7, was used to evaluate the performance of each neural network. A batch size of 16 was employed during training to accelerate network convergence; this reduced batch size also served as a form of regularization [79].

Initially, the 1D-CNN was trained for 300 epochs. However, this led to severe overfitting, with the network achieving 96.6% accuracy on the training set but only 84.2% on the validation set. To mitigate overfitting, dropout layers and pooling operations were incorporated into the network design, alongside the implementation of early stopping during training.

Determining the optimal point for early stopping presented a challenge. Analysis of the 1D-CNN learning curves over the initial 300 epochs (upper-left subfigure of Figure 14) revealed that training accuracy plateaued near 100% after approximately 200 epochs, indicating that continuing beyond this point could exacerbate overfitting [32]. Consequently, the early stopping threshold for the second training run was set at 200 epochs. As shown in the second subfigure of the 1D-CNN column in Figure 14, training was halted at this point, yielding 94.8% accuracy on the training set and 82.9% on the validation set. While overfitting was reduced compared to the initial 300-epoch training, a substantial accuracy gap of nearly 12% remained, warranting a third training attempt.

Observing the performance curve from the second experiment, the network achieved 100% validation accuracy at around 150 epochs. Thus, the early stopping point for the third training attempt was adjusted to 140 epochs, 10 epochs earlier. During this run, training ceased at 140 epochs, resulting in training and validation accuracies of 83.2% and 77.5%, respectively. Overfitting was significantly reduced to less than 6%, marking a substantial improvement over the approximate 12% gap observed in the previous attempts. These performance trends are illustrated in Figure 15. Notably, early stopping increased validation accuracy while decreasing training accuracy, and narrowed the accuracy gap between the training and validation sets, demonstrating its effectiveness in mitigating overfitting.

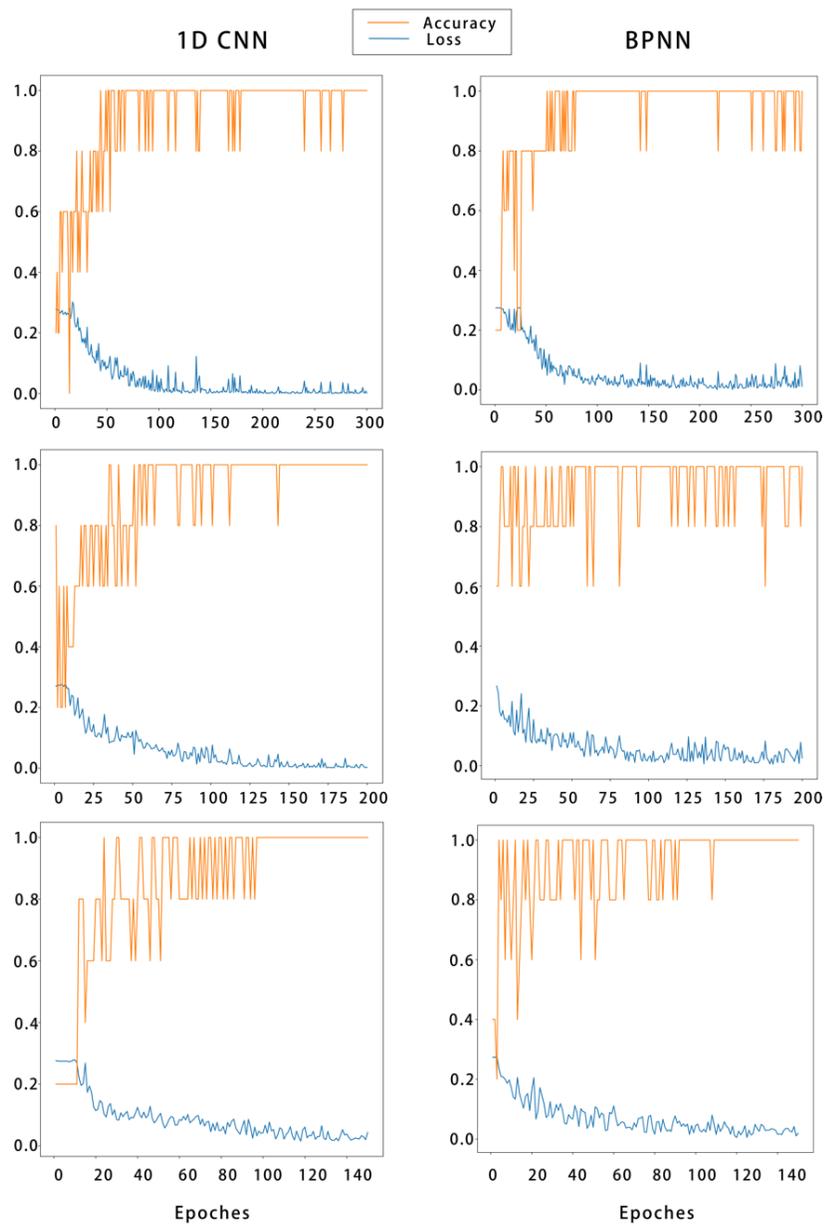

Figure 13. The NN learning curves.

The BPNN was trained using a similar procedure. Initially, training was conducted for 300 epochs. The BPNN learning curve, shown at the top row of Figure 14, indicates that the network achieves stability and near-perfect accuracy on the training set up to approximately 200 epochs. However, the training accuracy fluctuates markedly between 80% and 100%, suggesting that the network begins to memorize training samples and struggles to generalize to unseen data. Consequently, early stopping was implemented at 200 epochs during the second training round. The training set accuracy decreased by 2.3%, while the validation and test set accuracies dropped by only 0.3% and 0.2%, respectively. Despite the overall reduction in accuracy, the gap between training and validation/test accuracies narrowed by at least 2%, indicating a preliminary reduction in overfitting.

To further address overfitting, a third training round with early stopping was performed. Due to the difficulty in determining the optimal stopping epoch from the second learning curve, training was halted at 140 epochs—consistent with the early stopping point used for the 1D-CNN. The learning curve for this third round, shown in the lower right corner of Figure 14, demonstrates that the BPNN ceased training after reaching a stable 100% training accuracy. Figure 15 illustrates the BPNN's performance across datasets throughout these iterations. While training and validation accuracies decreased, the test set accuracy improved compared to the first training round, indicating that early stopping not only mitigates overfitting but also enhances generalization on unseen data.

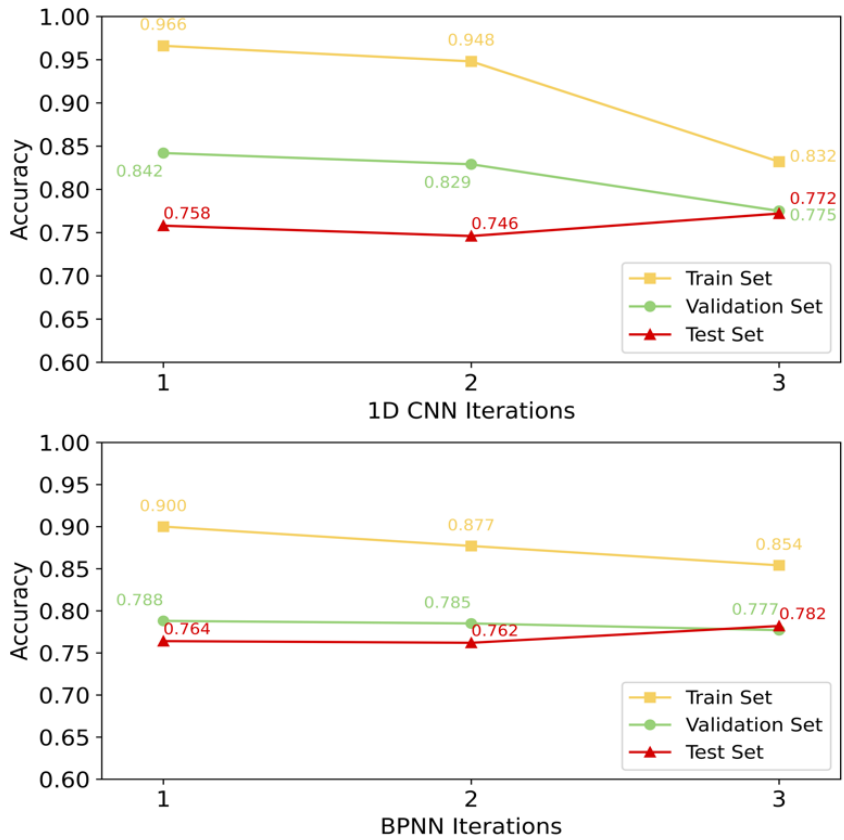

Figure 14. The iteration process of neural networks.

Overall, both neural network classifiers achieved at least 77% accuracy following the third round of early stopping. To provide a more detailed evaluation, confusion matrices were constructed to illustrate classification performance for each emotion category. Figure 16 presents these confusion matrices, where Figure 16(a), (c), and (e) correspond to the 1D-CNN model after the first, second, and third training rounds on the validation set, respectively. Figure 16(b), (d), and (f) show the BPNN model's confusion matrices from the same training rounds.

Comparison after 300 epochs reveals that the 1D-CNN outperforms the BPNN in accurately classifying negative emotions. Although the BPNN tends to misclassify bad emotions as neutral, both models frequently confuse Bad emotions with Good ones. Notably, the similarly shaded squares in Figure 16(c) and (d) indicate comparable error rates for both networks when misclassifying Good and Bad samples as Neutral. Furthermore, the darker purple hue at coordinate (G, G) in the 1D-CNN's confusion matrix compared to that of the BPNN indicates superior accuracy of the 1D-CNN in identifying Good emotions.

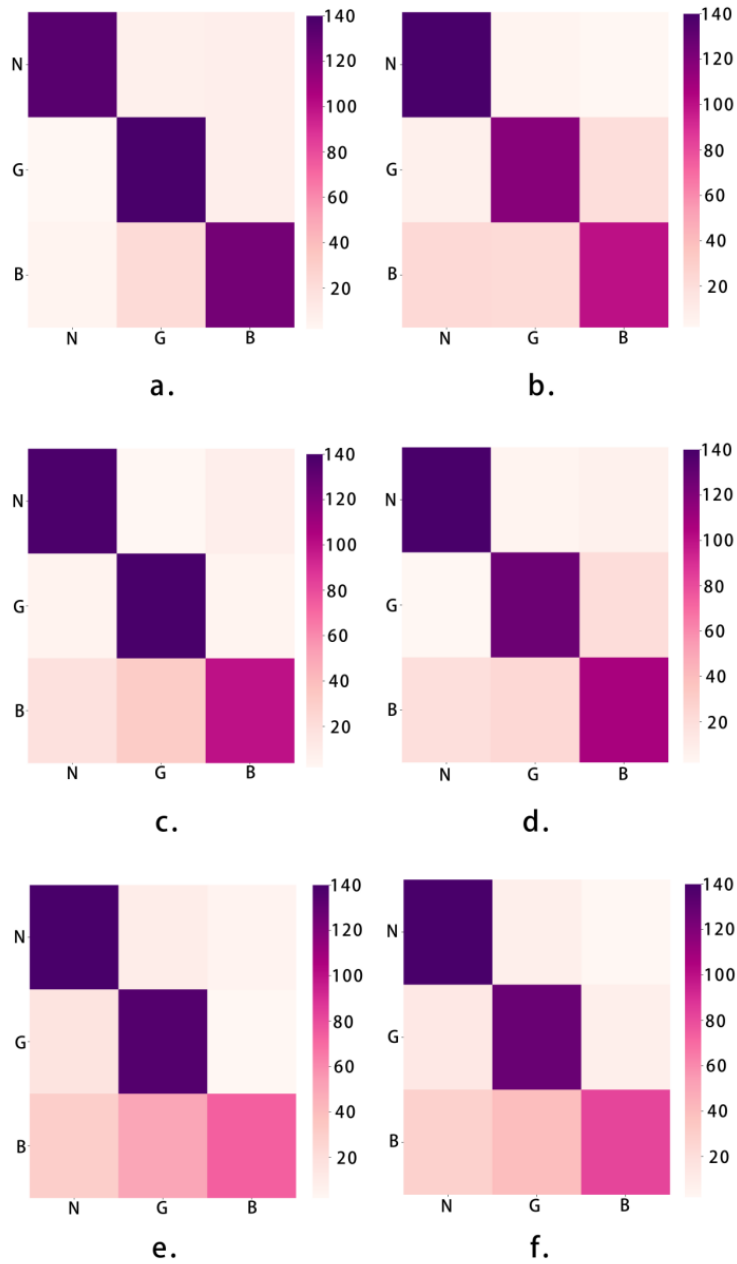

Figure 15. Confusion matrices of 1D-CNN (left column) and BPNN (right column). The vertical axis represents the true emotion labels, while the horizontal axis denotes the predicted labels.

Figures 16(e) and 16(f) illustrate the performance of the third round of early stopping training. Notably, the cell corresponding to (B, B) exhibits a pink hue, indicating that both neural networks continue to face challenges in accurately recognizing samples labeled as Bad emotion. Overall, the 1D-CNN trained for 300 epochs demonstrates superior performance on the validation set, as reflected in its confusion matrix. However, this does not suggest that early stopping was ineffective; poor validation performance does not necessarily imply poor generalization to unseen data. Rather, the networks trained with early stopping can be regarded as more generalizable, avoiding overfitting to the training and validation sets. This interpretation is supported by the observed increase in test set accuracy (represented by the red line) in Figure 15. Furthermore, Figure 15 indicates that the models obtained after the third early stopping round remain relatively weak learners, as their test set accuracies for both networks hover slightly below 80%.

In summary, while the SVM achieves just over 70% accuracy, it struggles notably with classifying Bad emotion samples. The neural networks attain approximately 80% accuracy overall but, according to the validation set confusion matrices, may exhibit particular difficulty with negative emotion classification. This limitation will be addressed in the subsequent ensemble learning strategy, which aims to combine these individual weak learners into a more robust classifier.

### 5.7 Establishing the ensemble learning model

Building the ensemble learning model involves training the meta learner using a stacking approach based on the outputs of the base learners. In stacking ensemble learning, the predictions from the base learners' validation sets serve as the training inputs for the meta learner. However, this approach has a limitation: the meta learner's training sample size is typically very small because the validation sets used by the base learners constitute only a fraction of the overall dataset. This makes partitioning the validation set for the meta learner challenging, necessitating the use of the LOOCV method as shown in Figure 7(c). This technique involves iteratively selecting one sample from the full dataset as the validation set while using the remaining N−1 samples for training. This process is repeated N times to ensure that each sample serves as the validation set exactly once. LOOCV is particularly well-suited for small datasets. In this study, an SVM combined with a grid search under LOOCV is employed to identify near-optimal hyperparameter values for C and γ of the meta learner.

Based on prior grid search experience with the base learners' SVMs, it is unlikely that optimal values of C and γ will exceed 3. Consequently, the search range for both parameters is restricted to the closed interval of [1,3]. The search converges at C=γ=0.1, forming the basis for constructing the meta learner. Once the meta learner is trained, the full ensemble learning model is assembled: each base learner first generates predictions on the data, which are then combined as new feature inputs to the meta learner. The meta learner subsequently produces the final emotion classification.

### 5.8 Analysis of results

The test set results presented in Table 11 demonstrate that the ensemble learning classifier achieves accuracy, recall, and F1 scores of 84% or higher across all emotion categories. Notably, the Neutral category exhibits the best classification performance, with all metrics reaching 88%. Similarly, the Bad emotion category maintains consistent performance, with all measures at 87%, representing a significant improvement over the base learner approach. Although the Good emotion category scores are slightly lower compared to the others, they remain at a robust 84% or above. Overall, these results validate the effectiveness of the ensemble learning strategy in enhancing classification accuracy across multiple emotion classes.

The ensemble learning model attains an impressive overall classification accuracy of 86%, reflecting an 8% to 17% improvement compared to individual base learners. These strong classification outcomes, coupled with balanced evaluation metrics, underscore the ensemble classifier's superior performance and robustness. It reliably classifies the majority of samples within each category, confirming its efficacy in sentiment recognition tasks.

A comparison of F1, Precision, and Recall metrics between the base learners and the ensemble model, depicted in Figure 17, reveals substantial gains in recognizing negative emotions. Both the F1 score and precision of the ensemble method exhibit marked improvements. While base learners may achieve higher recall or precision in certain categories, the upward trend observed in the F1 score subplot confirms that the ensemble learning approach presented here offers superior overall classification capability.

Table 11. Results of ensemble model on the test set.

| Emotion | Precision | Recall | F1 |
|---|---|---|---|
| G | 0.84 | 0.85 | 0.84 |
| N | 0.88 | 0.88 | 0.88 |
| B | 0.87 | 0.87 | 0.87 |
| Accuracy: 0.86 | | | |

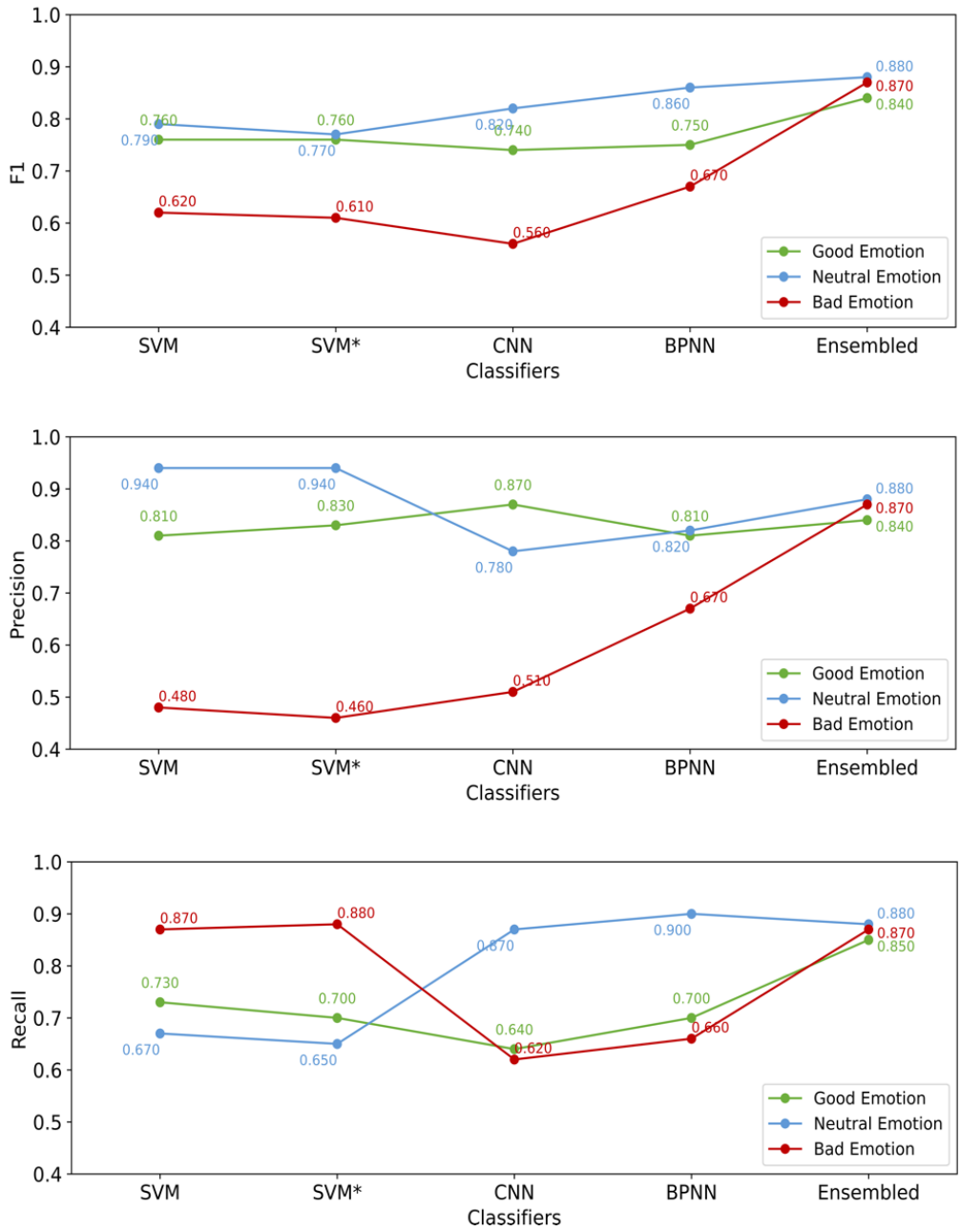

Figure 16. Classification results of the classifiers for each emotion. SVM denotes the final support vector machine obtained from lattice search using cross-validation, and SVM* denotes the final support vector machine obtained from lattice search using nested cross-validation.

## 6 CONCLUSION

This study presents a robust audio-only emotion recognition framework designed to classify movie scenes into three categories—Good, Neutral, and Bad—using an ensemble learning approach. Unlike multimodal or vision-based methods that require substantial computational resources, the proposed model relies exclusively on audio features, making it well-suited for deployment on devices with limited hardware capabilities. By integrating multiple weak learners, including SVMs and neural networks, within a stacking ensemble and combining this with a comprehensive data processing pipeline, the model achieves an overall accuracy of 86% on a real-world movie dataset—significantly outperforming individual base learners. Importantly, the model maintains balanced performance across all emotion classes, including the typically challenging Bad category. These results underscore the practical potential of high-performance, audio-based emotion recognition systems for home and mobile applications, offering a scalable solution for intelligent media recommendation, content filtering, and affective computing.

## ACKNOWLEDGMENTS

This research was partially supported by National Natural Science Foundation of China (52161042); Guangxi Science and Technology Major Program (2024AA29055); and the 100 Scholar Plan of the Guangxi Zhuang Autonomous Region of China (2018).